\documentclass[journal]{IEEEtran}

\usepackage{amsmath, amssymb, graphicx, upgreek, cite, marginnote, color}
\usepackage{epsfig}
\usepackage{enumitem}

\usepackage{algorithm}
\usepackage{algorithmic}
\usepackage{amsfonts}
\usepackage{graphicx}
\usepackage{textcomp}
\usepackage{multirow}
\usepackage{booktabs}   
\usepackage[T1]{fontenc}
\usepackage{mathptmx}

\usepackage[
    colorlinks=true,
    linkcolor=blue,
    citecolor=blue,
    urlcolor=blue
]{hyperref}

\newlength\myindent
\ifCLASSINFOpdf
\else
\fi
\begin{document}

\title{Manifold-Constrained PET Reconstruction with Learned Flow-Matching Priors}
\author{Hengjia Ran, \IEEEmembership{Student Member, IEEE}, Jie Luo, Yutao Zhu, Rui Hu, Fenglin Zhan, Jianyong Jiang, \\Huafeng Liu, \IEEEmembership{Senior Member, IEEE}, and Bo Zhao, \IEEEmembership{Member, IEEE}
	\thanks{This work was supported in part by the National Natural Science Foundation of China under Grant 62427807 and the Zhejiang Provincial ‘Jianbing Lingyan+X’ Science and Technology Program under Grant 2025C01127.}
	\thanks{H. Ran, Y. Zhu, and B. Zhao are with the ZJU-UIUC Institute, Zhejiang University, Haining 314400, China, and are also with the College of Information Science and Electronic Engineering, Zhejiang University, Hangzhou 310027, China.}
	\thanks{J. Luo is with the College of Optical Science and Engineering, Zhejiang University, Hangzhou 310027, China.}
	\thanks{R. Hu was with the College of Optical Science and Engineering, Zhejiang University, Hangzhou 310027, China. He is now with Vivo Mobile Communication Co., Ltd., Hangzhou 311100, China.}
	\thanks{F. Zhan is with the Department of Nuclear Medicine, the First Affiliated Hospital of the University of Science and Technology of China, Hefei 230001, China.}
	\thanks{J. Jiang is with the School of Physics and Astronomy and the Key Laboratory of Beam Technology of the Ministry of Education, Beijing Normal University, Beijing 100875, China. }
	\thanks{H. Liu is with the College of Biomedical Engineering and Instrument Science, Zhejiang University, Hangzhou 310027, China.}}

\maketitle

\begin{abstract}
Image reconstruction for positron emission tomography (PET) is an ill-posed Poisson inverse problem. In this work, we introduce an unsupervised, optimization-based reconstruction framework that employs a flow-matching generative model as a learned manifold prior. We train the generative model on high-quality PET images to learn a differentiable ordinary differential equation (ODE)-based transport from a Gaussian latent distribution to the PET image distribution. We incorporate this generator as an explicit manifold constraint into a regularized Poisson likelihood formulation. To solve the resulting problem, we develop an alternating direction method of multipliers (ADMM) algorithm that integrates an expectation-maximization (EM)-type surrogate update for Poisson data fidelity with gradient-based latent-space optimization. Numerical simulations and in vivo PET experiments demonstrate improved reconstruction accuracy compared with conventional reconstruction methods and recent deep learning-based approaches, particularly in low-count settings, with competitive computational efficiency.
\end{abstract}


\IEEEpeerreviewmaketitle

\section{Introduction}
\label{sec:introduction}
\IEEEPARstart{P}{ositron} emission tomography (PET) is a widely used medical imaging modality for visualizing and quantifying physiological and biochemical processes \textit{in vivo} \cite{cherry2012physics}.  By measuring the spatial distribution of radiotracers, PET enables the assessment of glucose metabolism, receptor binding, and perfusion for various clinical applications in neurology, oncology, and cardiology \cite{valk2006pet}. However, reconstructing high-quality PET images remains challenging due to the ill-posed nature of the inverse problem and the Poisson statistics of the measured data. These challenges become particularly pronounced in low-dose and short-acquisition protocols, where reduced photon counts often lead to increased statistical noise and reconstruction artifacts.

Statistical model-based iterative reconstruction has played a central role in PET image reconstruction for decades \cite{ollinger1997positron}. Maximum-likelihood (ML) reconstruction \cite{shepp1982maximum,hudson1994accelerated} explicitly models the Poisson statistics of PET measurements but is prone to noise amplification, often requiring early stopping in practice. Classical statistical reconstruction methods \cite{Levitan1987MAPEM,hebert1989gem,green1990bayesian,fessler1994penalized,DePierro1995, Zhou2007Wavelet,burger2014total,wang2015pet,chen2015sparse} address these limitations through smoothness, edge-preserving, and sparsity-promoting regularization, as well as kernel-based representations. Although these approaches improve noise characteristics, their limitations in capturing complex image statistics may lead to oversmoothing and loss of fine structural details.

Over the past decade, deep learning has emerged as a powerful paradigm for PET reconstruction \cite{Reader2021DeepLearning}. Supervised approaches, including end-to-end reconstruction \cite{haggstrom2019deeppet,hu2021dpir} and deep unrolling methods \cite{gong2019emnet,mehranian2020model,lim2020improved}, have achieved substantial improvements in reconstruction quality by leveraging paired training data. However, these approaches often require paired measurements and high-quality reference images, which can be difficult to obtain across different scanners, acquisition protocols, and radiotracers. Moreover, their performance may degrade under distribution shifts between training and deployment conditions.

Unsupervised learning offers an alternative to supervised reconstruction by exploiting implicit image priors or learning expressive priors from unpaired high-quality datasets. Training-data-free approaches, such as deep image priors (DIP)~\cite{gong2018pet} and certain implicit neural representation (INR) methods~\cite{moussaoui2025implicit}, use network architectures as implicit regularizers but often require computationally intensive optimization for individual reconstructions. More recently, diffusion-based generative models~\cite{ho2020ddpm,song2021score_sde} have shown promising results in PET reconstruction~\cite{singh2024score,Webber2025LikelihoodScheduled,Hashimoto2026DDIP}. These methods leverage learned image distributions to regularize reconstruction, often through iterative sampling involving numerous neural network evaluations, which can incur substantial computational costs.

Flow matching has recently emerged as a compelling generative modeling framework~\cite{lipman2023flow,liu2022flow}. By learning a time-dependent velocity field, it enables continuous transport from a simple base distribution to the data distribution through an ordinary differential equation (ODE), facilitating efficient deterministic image generation. The resulting differentiable mapping from the latent space to the image domain makes flow matching well suited for gradient-based optimization in imaging inverse problems.

In this paper, we introduce a novel unsupervised, optimization-based PET image reconstruction method that employs a flow-matching generative model as a learned manifold prior. We first train the generative model on high-quality PET images to capture the statistical characteristics of the underlying PET image distribution. We then formulate image reconstruction as a regularized Poisson likelihood optimization problem constrained to the learned manifold. To solve this problem, we develop an alternating direction method of multipliers (ADMM) algorithm~\cite{boyd2011admm} that integrates an expectation-maximization (EM)-type surrogate update for Poisson data fidelity with gradient-based latent-space optimization for manifold projection. Extensive experiments on simulated and in vivo PET data demonstrate improved reconstruction accuracy over conventional reconstruction methods and recent deep learning-based approaches, particularly in low-count settings, with competitive computational efficiency. A preliminary account of this work was presented in~\cite{Ran2026}, which, to our knowledge, was the first to investigate flow-matching generative models for PET reconstruction.


\section{Proposed Method}
\subsection{Problem Formulation}
PET aims to estimate the spatial distribution of radiotracer uptake $\mathbf{x}\in\mathbb{R}_+^{N}$ from measured sinogram data $\mathbf{y}\in\mathbb{Z}_+^{M}$. Due to the photon-counting nature of PET detectors, the measurement process can be modeled by 
\begin{equation}
	\mathbf{y}\sim \mathrm{Poisson}(\bar{\mathbf{y}}),
	\label{eq:pet_poisson_mean}
\end{equation}
where $ \mathbf{A}\in {\mathbb{R}}_{+}^{M\times N}$ denotes the system matrix, $\mathbf{r}\in {\mathbb{R}}_{+}^{M}$ denotes the expected background events (such as scatter and random coincidences), and $\bar{\mathbf y} = \mathbf A\mathbf x+\mathbf r \in \mathbb{R}_+^{M}$ denotes the mean of  $\mathbf y$. Given the independence across sinogram bins, the log-likelihood function can be written as
\begin{align}
	\log p(\mathbf{y}|\mathbf{x})
	=\sum_{i=1}^{M} \left(y_i \log(\bar{y}_i)-\bar{y}_i\right)-\sum_{i=1}^{M}\log(y_i!),
	\label{eq:loglik}
\end{align}
where $y_i$ and $\bar{y}_i$ denote the $i$th entries of $\mathbf{y}$ and $\bar{\mathbf{y}}$, respectively. Maximizing \eqref{eq:loglik} yields the ML estimate \cite{shepp1982maximum}. However, the ML estimate is prone to noise amplification, especially for low-count acquisitions, motivating the incorporation of prior information.

In this work, we employ a flow-matching generative model~\cite{lipman2023flow, liu2022flow} to learn a prior distribution of high-quality PET images. The model constructs a deterministic transport from a simple base distribution to the target image distribution through a neural ODE. Specifically, let $\mathbf{x}_0\in\mathbb{R}^{N}$ be a latent variable drawn from a standard multivariate Gaussian distribution $\mathcal{N}(\mathbf{0}, \mathbf{I})$. A time-dependent velocity field $\mathbf{v}_{\boldsymbol{\uptheta}}(\cdot, t)$, parameterized by $\boldsymbol{\uptheta}$, governs the following neural ODE:
\begin{equation}
	\frac{\mathrm{d}\mathbf{x}_t}{\mathrm{d}t}=\mathbf{v}_{\boldsymbol{\uptheta}}(\mathbf{x}_t,t),\qquad t\in[0,1],
	\label{eq:fm_ode}
\end{equation}
with $\mathbf{x}_{t=0}=\mathbf{x}_0$ and $\mathbf{x}_{t=1}=\mathbf{x}_1$. The terminal state $\mathbf{x}_1$ represents a generated PET image, defining a differentiable mapping $\mathbf{x}_1=\mathbf{G}_{\boldsymbol{\uptheta}}(\mathbf{x}_0)$.

Although PET images reside in a high-dimensional ambient space, their inherent geometric regularities suggest that the underlying image distribution is concentrated near a low-dimensional manifold. Flow matching provides a data-driven approach to capturing this structure by learning a generative mapping from a Gaussian latent distribution to the distribution of high-quality PET images. Specifically, the trained generator defines the generative image set
\begin{equation}
	\mathcal{M}_{\boldsymbol{\uptheta}}
	=\left\{
	\mathbf{G}_{\boldsymbol{\uptheta}}(\mathbf{x}_0)
	:\mathbf{x}_0\in\operatorname{supp}(p_0)
	\right\},
	\label{eq:generative_manifold}
\end{equation}
where $p_0$ denotes the standard Gaussian base distribution. We refer to $\mathcal{M}_{\boldsymbol{\uptheta}}$ as the learned generative image manifold.\footnote{While the range of a full-dimensional, invertible neural ODE need not be a low-dimensional manifold, the manifold interpretation mainly concerns concentration of the PET image distribution near lower-dimensional structure.} We incorporate the learned generative manifold as an explicit constraint in PET reconstruction to promote consistency with the statistical structure of high-quality PET images.

The time-dependent velocity field is learned using conditional flow matching (CFM)~\cite{lipman2023flow,liu2022flow}. Specifically, given a Gaussian latent sample $\mathbf{x}_0\sim\mathcal{N}(\mathbf{0},\mathbf{I})$ and a PET training image $\mathbf{x}_1\sim p_1$, we consider the linear interpolation path: 
\begin{equation}
	\mathbf{x}_t=(1-t)\mathbf{x}_0+t\mathbf{x}_1,
	\qquad t\in[0,1].
	\label{eq:cfm_linear_path}
\end{equation}
The corresponding conditional velocity field is $\mathbf{x}_1-\mathbf{x}_0$. The network parameters $\boldsymbol{\uptheta}$ are optimized by minimizing the following CFM loss:
\begin{equation}
	\mathcal{L}_{\mathrm{CFM}}(\boldsymbol{\uptheta})
	=\mathbb{E}_{t,\mathbf{x}_0,\mathbf{x}_1}
	\left[
	\left\|\mathbf{v}_{\boldsymbol{\uptheta}}(\mathbf{x}_t,t)
	-(\mathbf{x}_1-\mathbf{x}_0)\right\|_2^2
	\right],
	\label{eq:cfm_loss}
\end{equation}
where $t\sim\mathcal{U}[0,1]$ and $\mathbf{x}_0$ and $\mathbf{x}_1$ are sampled independently from their respective distributions. 

After training, generating an image amounts to solving the ODE in \eqref{eq:fm_ode} from $t=0$ to $t=1$:
\begin{equation}
	\mathbf{x}_1 = \mathbf{G}_{\boldsymbol{\uptheta}}(\mathbf{x}_0) = \mathbf{x}_0+\int_{0}^{1}\mathbf{v}_{\boldsymbol{\uptheta}}(\mathbf{x}_t,t)\,\mathrm{d}t.
	\label{eq:ode_solution}
\end{equation}
In practice, we can approximate \eqref{eq:ode_solution} using the forward Euler method: 
\begin{equation}
	\mathbf{x}_{t+\Delta t}=\mathbf{x}_{t}+\Delta t\,\mathbf{v}_{\boldsymbol{\uptheta}}(\mathbf{x}_t,t),
	\label{eq:euler}
\end{equation}
where $\Delta t = 1/L$ and $L$ denotes the number of Euler steps. 

Incorporating the learned generative manifold constraint and a Gaussian latent prior, we formulate PET reconstruction as the following regularized Poisson likelihood optimization problem:
\begin{align}
	\min_{\mathbf{x}\ge \mathbf{0},\,\mathbf{x}_0}\;\;&
	\sum_{i=1}^{M}\Big(\bar{y}_i-y_i\log(\bar{y}_i)\Big)
	+\frac{\mu}{2}\|\mathbf{x}_0\|_2^2,
	\nonumber\\
	\text{s.t.}\;\;&
	\mathbf{x}=\mathbf{G}_{\boldsymbol{\uptheta}}(\mathbf{x}_0),
	\label{eq:main_problem}
\end{align}
where $\mu>0$ controls the strength of Gaussian latent regularization, which favors latent representations in higher-density regions of the base distribution.\footnote{For
	the standard multivariate Gaussian base distribution, $p_0(\mathbf{x}_0)\propto\exp(-\|\mathbf{x}_0\|_2^2/2)$. Consequently, the quadratic latent penalty is proportional to the negative log-density of $p_0$, up to an additive constant.} Importantly, the Gaussian latent penalty provides an additional regularization mechanism beyond the generative constraint by discouraging latent representations in low-density regions of the base distribution~\cite{Ran2026_EMBC}.

\subsection{Solution Algorithm}
We develop an ADMM algorithm~\cite{boyd2011admm} to solve the constrained optimization problem in \eqref{eq:main_problem}. We first incorporate the nonnegativity constraint through an indicator function, yielding the following equivalent formulation:
\begin{align}
	\min_{\mathbf{x},\,\mathbf{x}_0}\;\;&
	\sum_{i=1}^{M}\Big(\bar{y}_i-y_i\log(\bar{y}_i)\Big)
	+\frac{\mu}{2}\|\mathbf{x}_0\|_2^2
	+\iota_{\mathbb{R}_+^N}(\mathbf{x}),
	\nonumber\\
	\text{s.t.}\;\;&
	\mathbf{x}=\mathbf{G}_{\boldsymbol{\uptheta}}(\mathbf{x}_0),
	\label{eq:main_problem_indicator}
\end{align}
where $\iota_{\mathbb{R}_+^N}(\mathbf{x})$ is the indicator function of the nonnegative orthant, defined by
\begin{equation}
	\iota_{\mathbb{R}_+^N}(\mathbf{x})=
	\begin{cases}
		0, & \mathbf{x}\geq\mathbf{0},\\
		+\infty, & \text{otherwise}.
	\end{cases}
	\label{eq:nonnegative_indicator}
\end{equation}

The augmented Lagrangian associated with \eqref{eq:main_problem_indicator}
can be written as
\begin{align}
	\mathcal{L}_{\rho}(\mathbf{x},\mathbf{x}_0,\boldsymbol{\nu})
	&={} \sum_{i=1}^{M}\Big(\bar{y}_i-y_i\log(\bar{y}_i)\Big)
	+\frac{\mu}{2}\|\mathbf{x}_0\|_2^2
	+\iota_{\mathbb{R}_+^N}(\mathbf{x}) \nonumber\\
	&+\left\langle\boldsymbol{\nu},
	\mathbf{x}-\mathbf{G}_{\boldsymbol{\uptheta}}(\mathbf{x}_0)
	\right\rangle  +\frac{\rho}{2}\left\|
	\mathbf{x}-\mathbf{G}_{\boldsymbol{\uptheta}}(\mathbf{x}_0)
	\right\|_2^2,
	\label{eq:augmented_lagrangian}
\end{align}
where $\boldsymbol{\nu}\in\mathbb{R}^N$ denotes the Lagrange multiplier and $\rho>0$ is the ADMM penalty parameter.

Applying the ADMM algorithm to \eqref{eq:augmented_lagrangian} yields the following alternating updates
\begin{align}
	\mathbf{x}^{n+1}
	&=\arg\min_{\mathbf{x}}\;\mathcal{L}_{\rho}(\mathbf{x},\mathbf{x}_0^{n},\boldsymbol{\nu}^{n}),
	\label{eq:admm_x}
	\\
	\mathbf{x}_0^{n+1}
	&=\arg\min_{\mathbf{x}_0}\;\mathcal{L}_{\rho}(\mathbf{x}^{n+1},\mathbf{x}_0,\boldsymbol{\nu}^{n}),
	\label{eq:admm_x0}
\end{align}
followed by the Lagrangian multiplier update
\begin{equation}
	\boldsymbol{\nu}^{n+1}
	=\boldsymbol{\nu}^{n}+\rho\left(\mathbf{x}^{n+1}-\mathbf{G}_{\boldsymbol{\uptheta}}(\mathbf{x}_0^{n+1})\right).
	\label{eq:admm_lambda}
\end{equation}

Next, we describe the solutions to the subproblems in
\eqref{eq:admm_x} and \eqref{eq:admm_x0} in detail.
The $\mathbf{x}$-subproblem in \eqref{eq:admm_x} can be written as
\begin{align}
	\mathbf{x}^{n+1}
	=\arg\min_{\mathbf{x}}\;\;&
	\sum_{i=1}^{M}\Big(\bar{y}_i-y_i\log(\bar{y}_i)\Big)
	+\iota_{\mathbb{R}_+^N}(\mathbf{x})\nonumber\\
	&+\frac{\rho}{2}\|\mathbf{x}-\mathbf{d}^n\|_2^2,
	\label{eq:x_subproblem_indicator}
\end{align}
where $\mathbf{d}^n=\mathbf{G}_{\boldsymbol{\uptheta}}(\mathbf{x}_0^n)
-\boldsymbol{\nu}^n/\rho$. This is equivalent to the following nonnegativity-constrained formulation: 
\begin{align}
	\mathbf{x}^{n+1}=\arg\min_{\mathbf{x}\ge \mathbf{0}}\;\;
	\sum_{i=1}^{M}\Big(\bar{y}_i-y_i\log(\bar{y}_i)\Big)
	+\frac{\rho}{2}\|\mathbf{x}-\mathbf{d}^n\|_2^2.
	\label{eq:x_subproblem}
\end{align}
This is a penalized maximum-likelihood reconstruction problem, which enforces Poisson data fidelity while keeping $\mathbf{x}$ close to the dual-corrected generator output. 

To solve \eqref{eq:x_subproblem} efficiently, we adopt an EM-style majorization approach \cite{DePierro1995,lange2000optimization} for the Poisson term. Specifically, an inner index $k$ is introduced with initialization $\mathbf{x}^{n,0}=\mathbf{x}^{n}$. For $k=0,1,\ldots,K-1$, we construct a surrogate
\begin{equation}
	Q(\mathbf{x};\mathbf{x}^{n,k})=-(\mathbf{b}^{n,k})^{T}\log\mathbf{x}+\mathbf{s}^{T}\mathbf{x},
	\label{eq:surrogate_Q}
\end{equation}
where $\mathbf{s}=\mathbf{A}^{T}\mathbf{1}$ and
\begin{equation}
	\mathbf{b}^{n,k}
	=\mathbf{x}^{n,k}\odot\left(\mathbf{A}^{T}\frac{\mathbf{y}}{\bar{\mathbf{y}}(\mathbf{x}^{n,k})}\right). 
	\label{eq:b_def}
\end{equation}
Here $\bar{\mathbf{y}}(\mathbf{x}^{n,k})=\mathbf{A}\mathbf{x}^{n,k}+\mathbf{r}$ denotes the Poisson mean predicted by $\mathbf{x}^{n,k}$, and $\odot$ denotes the element-wise multiplication. We then minimize the resulting objective function with the surrogate:
\begin{equation}
	\mathbf{x}^{n,k+1}
	=\arg\min_{\mathbf{x}\ge \mathbf{0}}\;
	Q(\mathbf{x};\mathbf{x}^{n,k})
	+\frac{\rho}{2}\|\mathbf{x}-\mathbf{d}^{n}\|_2^2.
	\label{eq:mm_update}
\end{equation}
The objective function in \eqref{eq:mm_update} is convex and separable across voxels. For positive voxel values, its first-order optimality condition is given by
\begin{equation}
	-\frac{\mathbf{b}^{n,k}}{\mathbf{x}}
	+\mathbf{s}
	+\rho(\mathbf{x}-\mathbf{d}^{n})
	=\mathbf{0}.
	\label{eq:vector_optimality}
\end{equation}
Multiplying \eqref{eq:vector_optimality} element-wise by $\mathbf{x}$ and
taking the nonnegative root gives the EM-type closed-form update:
\begin{equation}
	\mathbf{x}^{n,k+1}
	=
	\frac{
		-(\mathbf{s}-\rho \mathbf{d}^{n})+
		\sqrt{(\mathbf{s}-\rho \mathbf{d}^{n})^{2}+4\rho\,\mathbf{b}^{n,k}}
	}{2\rho},
	\label{eq:closed_form}
\end{equation}
with all operations applied element-wise. Note that the above closed-form update also accommodates zero-valued voxels, with vanishing logarithmic coefficients interpreted using the standard limiting convention. After $K$ inner iterations, we set
\begin{equation}
	\mathbf{x}^{n+1}=\mathbf{x}^{n,K}.
	\label{eq:x_update}
\end{equation}
This update inherently preserves nonnegativity, and using a small $K$ provides an efficient update.

The $\mathbf{x}_0$-subproblem in \eqref{eq:admm_x0} is a regularized latent projection problem:
\begin{equation}
	\mathbf{x}_0^{n+1}
	=
	\arg\min_{\mathbf{x}_0}\;
	\left\|
	\mathbf{G}_{\boldsymbol{\theta}}(\mathbf{x}_0)-
	\left(\mathbf{x}^{n+1}+\frac{\boldsymbol{\nu}^{n}}{\rho}\right)
	\right\|_2^2
	+\lambda\|\mathbf{x}_0\|_2^2,
	\label{eq:latent_projection_alpha}
\end{equation}
where $\lambda=\mu/\rho$. This is a smooth nonlinear optimization problem, which we solve by a quasi-Newton algorithm, i.e., the limited-memory Broyden–Fletcher–Goldfarb–Shanno (L-BFGS) algorithm \cite{nocedal2006numerical}. Here we calculate the gradient by automatic differentiation. 

To improve computational efficiency, we adopt an inexact ADMM strategy, approximately solving the two subproblems in \eqref{eq:admm_x} and \eqref{eq:admm_x0} using a limited number of EM-type and L-BFGS iterations, respectively. We further employ warm-start initialization for both subproblems, using the solutions from the previous ADMM iteration as the initial estimates for the current iteration. 

\begin{figure*}[t]
	\centering
	\includegraphics[width=0.95\linewidth]{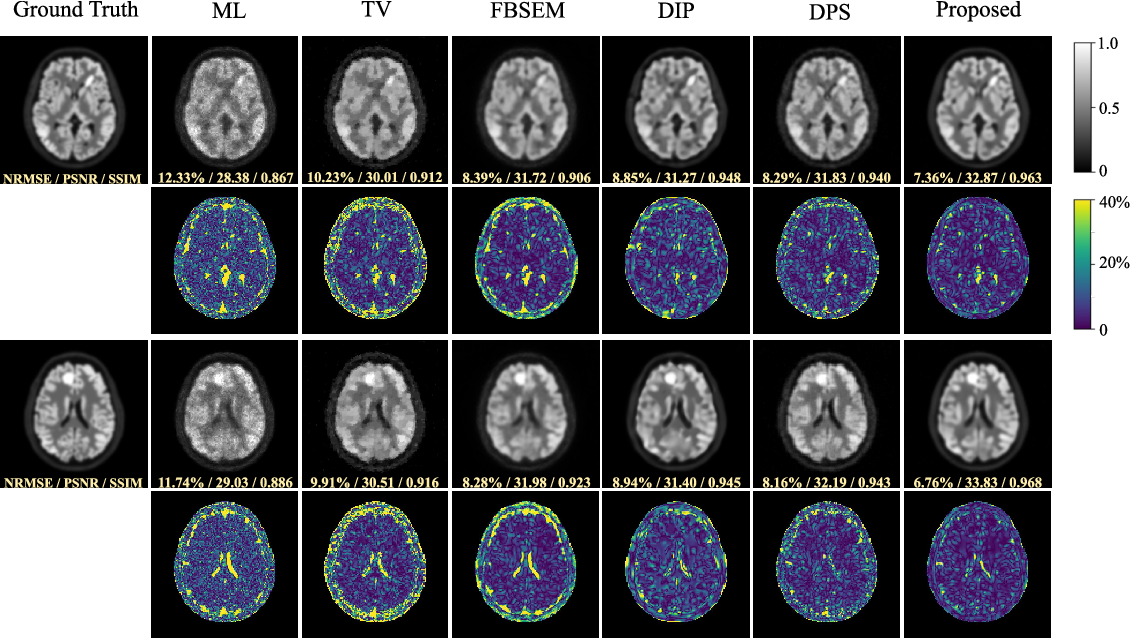}
	\caption{Comparison of reconstruction methods at the 50\% dose level in the BrainWeb simulation study. Two representative axial slices are shown, with the ground-truth and reconstructed images in the first and third rows and the corresponding relative error maps in the second and fourth rows. NRMSE, PSNR (dB), and SSIM are reported below each reconstruction.}
	\label{fig:fig1}
\end{figure*}

\section{EXPERIMENTS AND RESULTS}
\label{sec:guidelines}
\subsection{Experimental Setup}
\subsubsection{Simulation study}
We performed numerical simulations with known ground-truth activity distributions to evaluate the proposed method. Specifically, we conducted a 2D simulation study using BrainWeb digital phantoms \cite{AubertBroche2006phantoms} and a 3D simulation study using clinical images from the 2025 Ultra-Low Dose PET Imaging Challenge \cite{xue2025udpet}.

For the 2D simulation study, we used 20 anatomically realistic BrainWeb digital phantoms. The subjects were divided into 18 training cases, 1 validation case, and 1 independent test case. Simulated PET activity distributions were generated from these phantoms, and each volume was augmented using three random deformation realizations following \cite{Schramm2021brainwebPETMR}. Each axial image had a matrix size of $128\times128$ and an in-plane voxel size of $1.65\times1.65$ mm$^2$. The training data contained no tumor lesions. Tumors were manually introduced into the test phantom to evaluate reconstruction performance on pathological uptake patterns absent from the training data.
\begin{table}[t]
	\centering
	\caption{Quantitative comparison of reconstruction methods at five dose levels, averaged over the middle 20 axial slices of the BrainWeb test subject. The best result for each metric at each dose level is highlighted in \textbf{bold} and the second-best result is \underline{underlined}.}
	\label{tab:quantitative_comparison}
	\scriptsize
	\setlength{\tabcolsep}{2.5pt}
	\renewcommand{\arraystretch}{0.98}
	\begin{tabular}{@{} l c cccccc @{}}
		\toprule
		\multirow{2}{*}{\textbf{Metric}} & \multirow{2}{*}{\textbf{Dose}} & \multicolumn{6}{c}{\textbf{Method}} \\
		\cmidrule(lr){3-8}
		& & \textbf{ML} & \textbf{TV} & \textbf{FBSEM} & \textbf{DIP} & \textbf{DPS} & \textbf{Proposed} \\
		\midrule
		
		\multirow{5}{*}{\textbf{NRMSE} ($\downarrow$)}
		& 2\%  & 24.77\% & 19.53\% & 16.91\% & 19.68\% & \underline{16.24\%} & \textbf{15.36\%} \\
		& 5\%  & 20.55\% & 16.92\% & 13.74\% & 17.06\% & \underline{13.64\%} & \textbf{12.73\%} \\
		& 10\% & 17.84\% & 14.84\% & 12.72\% & 14.35\% & \underline{12.09\%} & \textbf{10.84\%} \\
		& 25\% & 14.41\% & 12.27\% & 10.01\% & 11.25\% & \underline{9.97\%}  & \textbf{8.55\%} \\
		& 50\% & 12.21\% & 10.31\% & \underline{8.29\%} & 9.51\% & 8.33\% & \textbf{7.09\%} \\
		\midrule
		
		\multirow{5}{*}{\textbf{PSNR} ($\uparrow$)}
		& 2\%  & 22.44 & 24.22 & 25.94 & 24.16 & \underline{26.11} & \textbf{26.84} \\
		& 5\%  & 24.06 & 25.91 & 27.56 & 25.80 & \underline{27.62} & \textbf{28.26} \\
		& 10\% & 25.29 & 26.86 & 28.22 & 27.19 & \underline{28.67} & \textbf{29.64} \\
		& 25\% & 27.15 & 28.63 & 30.31 & 29.30 & \underline{30.35} & \textbf{31.68} \\
		& 50\% & 28.58 & 30.06 & \underline{31.94} & 30.77 & 31.89 & \textbf{33.29} \\
		\midrule
		
		\multirow{5}{*}{\textbf{SSIM} ($\uparrow$)}
		& 2\%  & 0.402 & 0.724 & 0.797 & 0.734 & \underline{0.807} & \textbf{0.844} \\
		& 5\%  & 0.538 & 0.779 & 0.844 & 0.814 & \underline{0.860} & \textbf{0.895} \\
		& 10\% & 0.652 & 0.816 & 0.866 & 0.871 & \underline{0.890} & \textbf{0.925} \\
		& 25\% & 0.764 & 0.868 & 0.908 & \underline{0.922} & 0.920 & \textbf{0.948} \\
		& 50\% & 0.835 & 0.903 & 0.916 & \underline{0.940} & 0.937 & \textbf{0.966} \\
		\bottomrule
	\end{tabular}
\end{table}
The PET acquisition geometry was modeled after the Siemens Biograph Vision scanner, and the 2D PET forward model was implemented using the ParallelProj package \cite{Schramm2023parallelproj}, with each axial slice projected independently. To model the finite spatial resolution of the imaging system, a spatially invariant Gaussian point spread function (PSF) with a full width at half maximum (FWHM) of 4.5 mm was applied to the activity distributions before forward projection.

Simplified water-equivalent attenuation maps were generated by thresholding smoothed PET activity images to obtain support masks, which were subsequently smoothed and scaled by the linear attenuation coefficient of water, i.e., $0.0096$~mm$^{-1}$ \cite{Rausch2021}. Scatter contributions were estimated from the activity images and attenuation maps using the openSSS single-scatter simulation method \cite{santo2025opensss}, whereas random coincidence contributions were estimated from simulated delayed coincidence events \cite{brasse2005randoms}.

For the 2D BrainWeb simulations, the full-dose count scale was calibrated against a 3D reference acquisition with approximately $4.81\times10^8$ expected primary counts. This calibration was incorporated into the effective 2D acquisition model, yielding slice-dependent expected count levels. Low-dose PET data were simulated by randomly retaining the prescribed fraction of the full-count simulated events, with the retained count fraction used to represent the relative dose level. 

For the 3D simulation study, we used full-count reconstructed PET images from the 2025 Ultra-Low Dose PET Imaging Challenge \cite{xue2025udpet}. The dataset contains images acquired using two total-body PET scanners: a Siemens Biograph Vision Quadra scanner at the University of Bern, Switzerland, and a uEXPLORER scanner from United Imaging Healthcare, Shanghai, China. The Siemens and uEXPLORER datasets contain PET scans from 372 and 301 subjects, respectively.

The Siemens dataset was divided at the patient level into training, validation, and test sets in an 8:1:1 ratio. The flow-matching prior was trained using the Siemens training images and evaluated on the held-out Siemens test data. To assess generalization across scanner-specific image distributions, we additionally trained the generative priors using the United Imaging uEXPLORER images and evaluated them on simulations based on the same held-out Siemens test images, using the same Siemens forward model. Only the brain region was used, and all PET volumes were cropped to $128\times128\times80$ voxels for model training and evaluation.

The full-count clinical images were treated as ground-truth activity distributions for generating synthetic projection-domain measurements. The 3D PET forward model was implemented using the ParallelProj package \cite{Schramm2023parallelproj} based on the Siemens Biograph Vision scanner geometry. Simplified water-equivalent attenuation maps were constructed using the same procedure as in the 2D study. Scatter contributions were estimated from the activity images and attenuation maps using openSSS \cite{santo2025opensss}, whereas random coincidence contributions were estimated from simulated delayed coincidence events \cite{brasse2005randoms}. The full-dose acquisition was calibrated to approximately $4.81\times10^8$ expected primary counts per volume. Low-dose sinograms were generated by downsampling events from the full-count simulated measurements to the prescribed count levels.

\subsubsection{In vivo study}
We further evaluated the proposed method using in vivo PET data from 15 subjects who underwent clinical $[^{18}\mathrm{F}]$FDG PET scans on a Siemens Biograph Vision 8-ring PET/CT scanner at the First Affiliated Hospital of the University of Science and Technology of China. The study was approved by the local institutional review board, and all subjects provided informed consent. The subjects had a mean age of $54.3\pm16.9$ years, a mean body weight of $65.4\pm10.8$ kg, and a mean injected activity of $275.7\pm17.7$ MBq. Each scan had an acquisition duration of 600 s.

The measured list-mode events were rebinned into sinograms using the corresponding PET acquisition geometry. Attenuation correction was performed using subject-specific CT-based attenuation maps resampled and spatially aligned to the PET reconstruction grid. Random coincidence contributions were estimated from smoothed delayed-coincidence sinograms \cite{brasse2005randoms}, whereas scatter distributions were estimated using openSSS \cite{santo2025opensss}.

The full-count data were reconstructed using ordered-subsets expectation maximization (OSEM) \cite{hudson1994accelerated} with three iterations and five subsets, and the resulting images served as references. Low-count measurements were generated by randomly retaining 10\% of the full-count measured events. The flow-matching prior trained on the Siemens challenge training data for the 3D simulation study was applied to the low-count clinical sinograms without retraining on this independent dataset.
\begin{figure}[t]
	\centering
	\includegraphics[width=0.85\linewidth]{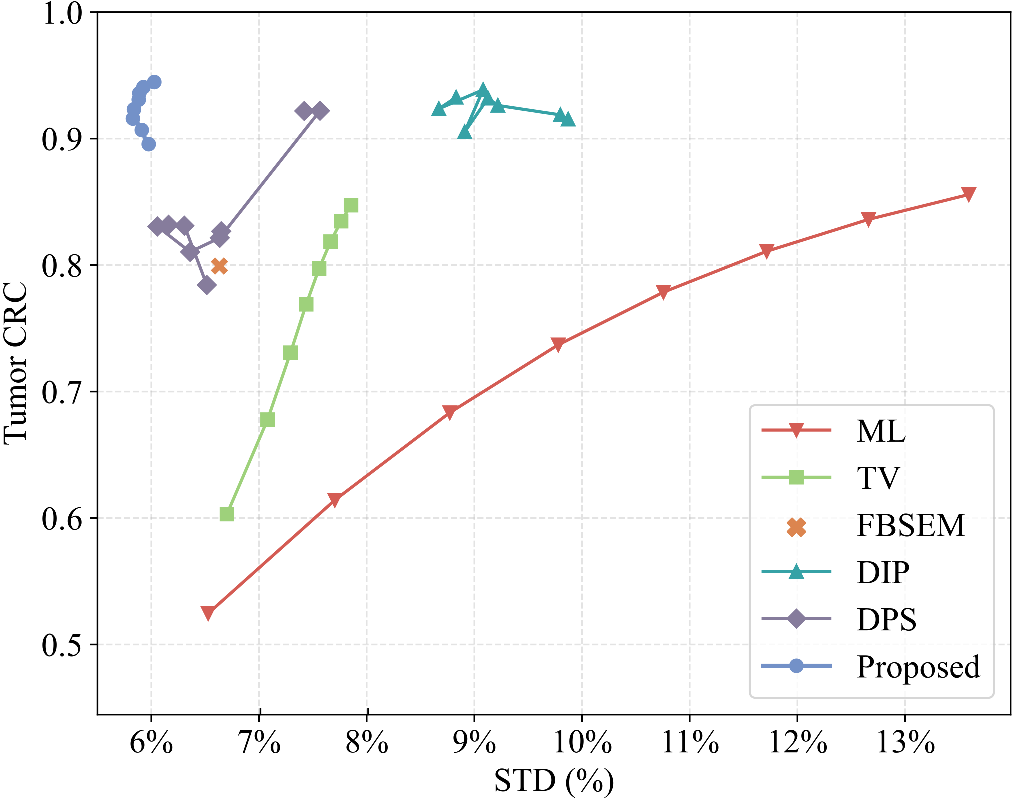} %
	\caption{Trade-off between tumor CRC and background STD in white-matter regions, evaluated over 10 independent noise realizations. The curves correspond to 15–50 iterations for ML, 1000–3100 iterations for TV, 5000–5400 iterations for DIP, data-consistency guidance weights ranging from 6 to 20 for DPS, and 6–15 iterations for the proposed method. FBSEM is shown as a single operating point corresponding to the output of the trained unrolled network.
	}
	\label{crc_std}
\end{figure}

\begin{figure*}[t]
	\centering
	\includegraphics[width = \linewidth]{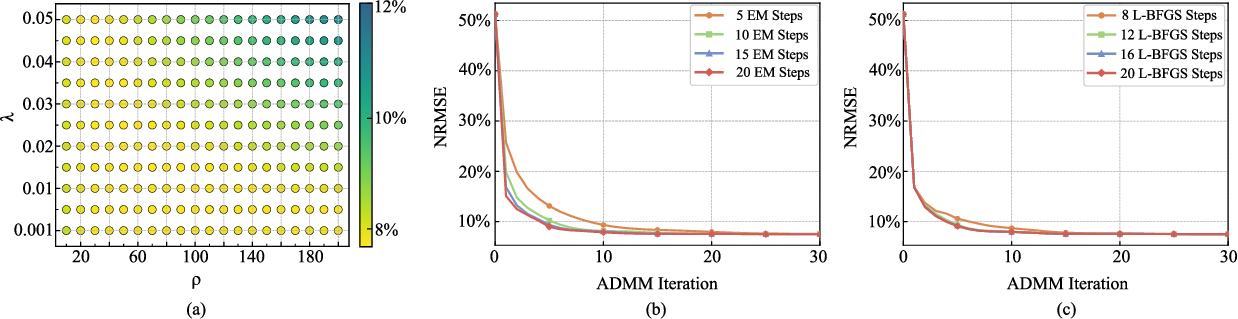} %
	\caption{Empirical convergence and parameter sensitivity of the proposed algorithm. (a) Final NRMSE for different values of the ADMM penalty parameter $\rho$ and latent regularization parameter $\lambda$. (b) NRMSE versus ADMM iteration for different numbers of EM-type updates in the Poisson data consistency subproblem. (c) NRMSE versus ADMM iteration for different numbers of L-BFGS iterations in the latent projection subproblem.}
	\label{robustness}
\end{figure*}

\begin{figure}[t]
	\centering
	\includegraphics[width=0.78\linewidth]{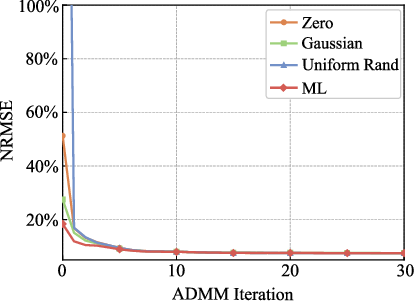} %
	\caption{Sensitivity of the proposed algorithm to initialization. NRMSE is plotted against ADMM iteration for zero, Gaussian random, uniform random, and ML-based initialization. It is clear that all four strategies approach nearly identical final NRMSE values.}
	\label{log-likelihood}
\end{figure}

\subsubsection{Method implementation}
The proposed method was implemented in Python 3.10 using PyTorch 2.8.0 on a Linux server equipped with an NVIDIA H100 NVL GPU with 94 GB of HBM3 memory. For the BrainWeb simulation study, the velocity field $\mathbf{v}_{\boldsymbol{\uptheta}}(\mathbf{x}_t,t)$ was parameterized by a residual U-Net \cite{ronneberger2015unet,he2016resnet} with a single axial slice as input and the corresponding single-slice velocity field as output. The network comprised four encoder stages and four decoder stages connected by skip connections. Each residual block contained two $3\times3$ convolutional layers, each followed by batch normalization and a rectified linear unit (ReLU). A $1\times1$ convolution was used in the shortcut connection when the channel dimensions differed. The base feature width was 64, with encoder feature widths of 64, 128, 256, and 512.

For the clinical-image-based simulation and in vivo studies, the same network architecture was used, with the number of channels increased from one to five at both the input and output. The network took a stack of five adjacent axial slices as input and predicted the corresponding five-slice velocity field, thereby incorporating inter-slice information. Both networks were trained using the Adam optimizer \cite{kingma2014adam} with an initial learning rate of $1\times10^{-4}$. 
\begin{figure}[t]
	\centering
	\includegraphics[width=0.80\linewidth]{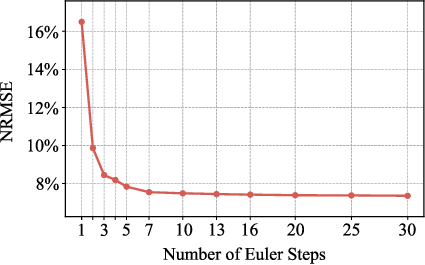} %
	\caption{Effect of the number of Euler steps in the flow-matching ODE solver on reconstruction NRMSE. The NRMSE decreases sharply up to approximately 10 steps, with only marginal improvements thereafter.}
	\label{Euler}
\end{figure}
\subsubsection{Method evaluation}
We compared the proposed method with five reconstruction methods: maximum-likelihood (ML) reconstruction \cite{shepp1982maximum} and total variation (TV)-regularized reconstruction \cite{burger2014total}, representing conventional model-based methods; a supervised deep-unrolling method (FBSEM) \cite{mehranian2020model}; and two unsupervised learning-based methods, namely deep image prior (DIP)-based reconstruction \cite{gong2018pet, Ulyanov2018DIP} and score-based diffusion posterior sampling (DPS) \cite{chung2022diffusion}.

To facilitate a controlled comparison, the DIP reconstruction network and the DPS score network used the same residual U-Net backbone as the proposed method's velocity-field network for each study. The learned reconstruction modules in FBSEM also used this backbone, but with a reduced base feature width to accommodate GPU memory constraints during training. All reconstruction methods were implemented on the same computational platform, and their hyperparameters were tuned using a separate validation dataset.

To assess the performance of reconstruction methods, we used the following quantitative metrics: normalized root mean square error (NRMSE), peak signal-to-noise ratio (PSNR), and structural similarity index (SSIM) \cite{Wang2004SSIM}. For region-of-interest (ROI) analysis, we also measured the contrast recovery coefficient (CRC), normalized inter-realization background standard deviation (STD), and within-ROI background coefficient of variation (CoV), following previous work \cite{wang2015pet,Hashimoto2026DDIP}. Here, CRC quantifies tumor contrast recovery, whereas STD and CoV quantify inter-realization and within-ROI background variability, respectively.

\begin{figure*}[t]
	\centering
	\includegraphics[width=\linewidth]{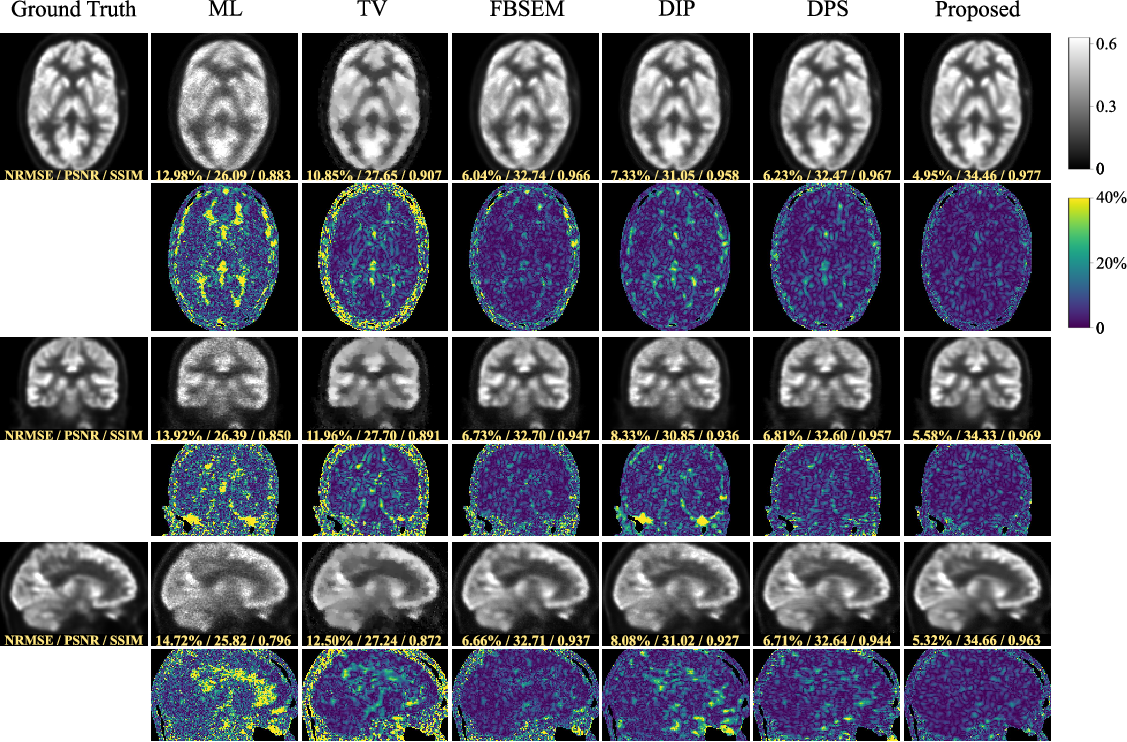}
	\caption{Comparison of reconstruction methods at the 25\% dose level in the clinical-image-based simulation study. Representative axial, coronal, and sagittal views are shown from top to bottom, with the ground truth and reconstructed images followed by their corresponding relative error maps. NRMSE, PSNR (dB), and SSIM are reported below each reconstruction.}
	\label{real_data}
\end{figure*}
\begin{table}[t] 
	\centering
	\caption{Quantitative comparison of reconstruction methods at five dose levels, averaged over 20 subjects in the clinical-image-based simulation study. The best result for each metric at each dose level is highlighted in \textbf{bold}, and the second-best result is \underline{underlined}.}
	\label{tab:patient_quant}
	\scriptsize
	\setlength{\tabcolsep}{2.5pt}
	\renewcommand{\arraystretch}{0.98}
	\begin{tabular}{@{} l c cccccc @{}}
		\toprule
		\multirow{2}{*}{\textbf{Metric}} & \multirow{2}{*}{\textbf{Dose}} & \multicolumn{6}{c}{\textbf{Method}} \\
		\cmidrule(lr){3-8}
		& & \textbf{ML} & \textbf{TV} & \textbf{FBSEM} & \textbf{DIP} & \textbf{DPS} & \textbf{Proposed} \\
		\midrule
		
		\multirow{5}{*}{\textbf{NRMSE} ($\downarrow$)}
		& 2\%  & 21.69\% & 17.27\% & \underline{11.38\%} & 17.08\% & 11.51\% & \textbf{8.80\%} \\
		& 5\%  & 17.50\% & 13.23\% & 9.70\% & 12.43\% & \underline{9.14\%} & \textbf{7.26\%} \\
		& 10\% & 15.00\% & 11.18\% & \underline{7.46\%} & 9.76\% & 7.73\% & \textbf{6.27\%} \\
		& 25\% & 12.44\% & 10.29\% & 6.15\% & 7.06\% & \underline{6.09\%} & \textbf{5.16\%} \\
		& 50\% & 10.19\% & 8.44\% & 5.30\% & 5.73\% & \underline{5.25\%} & \textbf{4.51\%} \\
		\midrule
		
		\multirow{5}{*}{\textbf{PSNR} ($\uparrow$)}
		& 2\%  & 25.97 & 27.95 & \underline{31.57} & 28.05 & 31.47 & \textbf{33.81} \\
		& 5\%  & 27.84 & 30.27 & 32.96 & 30.81 & \underline{33.48} & \textbf{35.48} \\
		& 10\% & 29.18 & 31.73 & \underline{35.25} & 32.91 & 34.93 & \textbf{36.75} \\
		& 25\% & 30.80 & 32.50 & 36.92 & 35.73 & \underline{37.01} & \textbf{38.45} \\
		& 50\% & 32.54 & 34.52 & 38.21 & 37.54 & \underline{38.30} & \textbf{39.62} \\
		\midrule
		
		\multirow{5}{*}{\textbf{SSIM} ($\uparrow$)}
		& 2\%  & 0.790 & 0.850 & 0.898 & 0.845 & \underline{0.918} & \textbf{0.955} \\
		& 5\%  & 0.843 & 0.888 & 0.918 & 0.905 & \underline{0.943} & \textbf{0.965} \\
		& 10\% & 0.875 & 0.905 & 0.951 & 0.936 & \underline{0.956} & \textbf{0.971} \\
		& 25\% & 0.903 & 0.921 & 0.962 & 0.963 & \underline{0.971} & \textbf{0.976} \\
		& 50\% & 0.931 & 0.944 & 0.973 & 0.975 & \underline{0.978} & \textbf{0.982} \\
		\bottomrule
	\end{tabular}
\end{table}
\subsection{Results}
\subsubsection{Simulation study}
Fig.~\ref{fig:fig1} shows reconstructions from ML, TV, FBSEM, DIP, DPS, and the proposed method at the 50\% dose level. The ML reconstruction is heavily corrupted by noise and streaking artifacts. Although the classical model-based method (TV) improves upon ML, it introduces oversmoothing and blurs fine anatomical structures. The learning-based approaches, including the supervised method (FBSEM) and the unsupervised methods (DIP and DPS), further improve reconstruction quality over the conventional methods. Among all methods, the proposed method achieves the best overall performance in terms of visual quality and quantitative metrics. For a more comprehensive quantitative assessment, we also compared all methods on the middle 20 axial slices of the test subject across five dose levels (2\%, 5\%, 10\%, 25\%, and 50\%). The results, reported in Table~\ref{tab:quantitative_comparison}, show that the proposed method achieves the lowest NRMSE and the highest PSNR and SSIM at all evaluated dose levels, consistent with the visual observations.

To evaluate generalization to unseen lesions, we inserted lesions into the test images while keeping all training images lesion-free. Reconstruction performance was evaluated over 10 independent noise realizations using tumor CRC and background STD to assess lesion contrast recovery and background variability, respectively. Fig.~\ref{crc_std} shows the CRC–STD trade-off for the different reconstruction methods. The unsupervised methods (DIP and DPS) yield higher CRC values than the supervised method (FBSEM). The proposed method achieves the highest CRC while maintaining substantially lower STD levels than the competing methods, indicating a more favorable trade-off between lesion contrast recovery and noise suppression. These results illustrate its ability to generalize to lesions absent from the training data.

\begin{figure*}[t]
	\centering
	\includegraphics[width=0.90\linewidth]{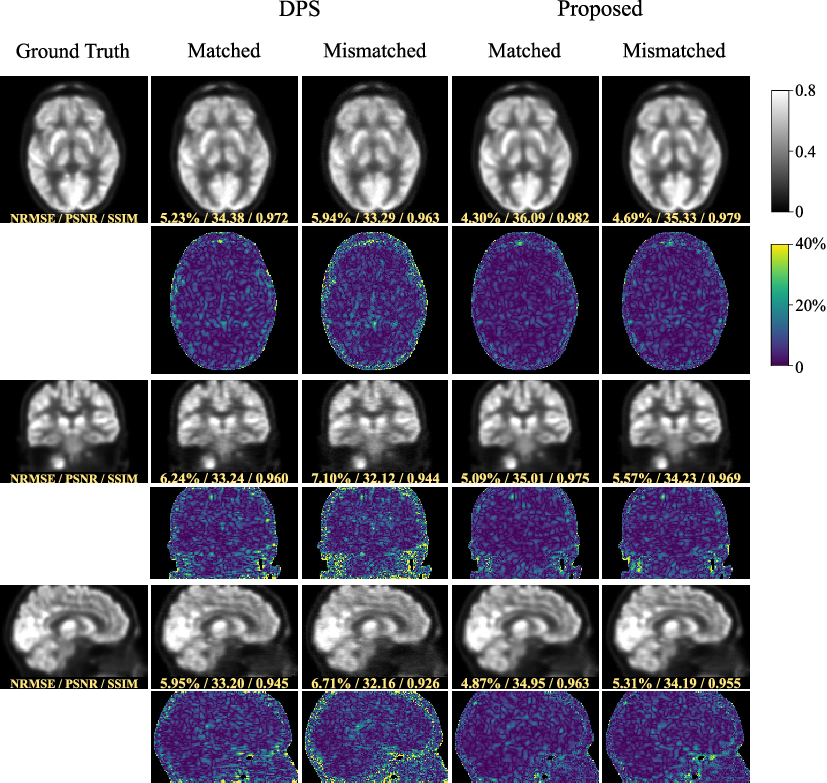}
	\caption{Cross-scanner generalization evaluation using clinical-image-based simulated data. DPS and the proposed method are compared under matched (Siemens training / Siemens testing) and mismatched (United Imaging training / Siemens testing) settings. Representative axial, coronal, and sagittal views are shown from top to bottom, with corresponding relative error maps below each reconstruction. NRMSE, PSNR (dB), and SSIM are reported below each reconstruction. The proposed method exhibits less performance degradation under scanner mismatch than DPS.}
	\label{cross_sanner}
\end{figure*}
\begin{figure*}[t]
	\centering
	\includegraphics[width=0.93\linewidth]{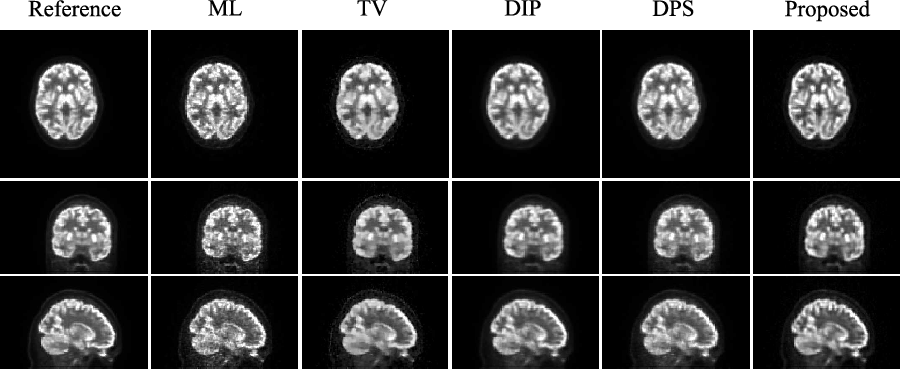}
	\caption{Comparison of reconstruction methods using in vivo PET data at the 10\% count-level from an independent patient dataset. Representative axial, coronal, and sagittal views are shown from top to bottom. The full-count reconstruction is provided as a reference.}
	\label{real}
\end{figure*}

\begin{figure}[t]
	\centering
	\includegraphics[width=0.85\linewidth]{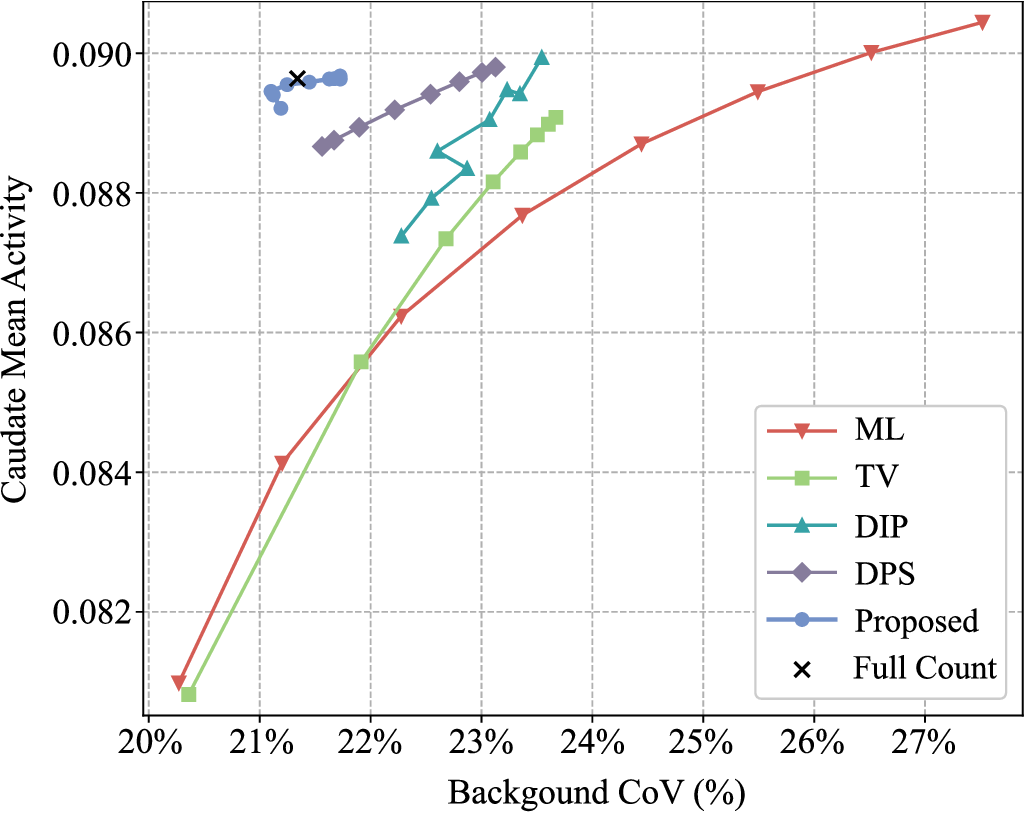} %
	\caption{
		Trade-off between mean caudate activity and background CoV in white-matter regions using 10\% count-level in vivo PET data from 15 subjects. The full-count reconstruction is marked as the reference. The curves correspond to 9–30 iterations for ML, 500–4000 iterations for TV, 1800–2500 iterations for DIP, data-consistency guidance weights ranging from 10 to 45 for DPS, and 10–30 iterations for the proposed method.
	}
	\label{cov}
\end{figure}

Fig.~\ref{robustness} evaluates the empirical convergence behavior and parameter sensitivity of the proposed algorithm.  In Fig.~\ref{robustness}(a), the final NRMSE varies moderately across a wide range of the ADMM penalty parameter $\rho$ and the latent regularization parameter $\lambda$. In Fig.~\ref{robustness}(b), using 5–20 EM-type updates in the Poisson data consistency subproblem yields similar final NRMSE values, although fewer updates lead to slower initial convergence. In Fig.~\ref{robustness}(c), using 8–20 L-BFGS iterations in the latent projection subproblem also yields similar convergence behavior, with nearly overlapping curves for 16 and 20 iterations. These results suggest diminishing improvements from additional inner iterations. With 15 EM-type updates and 12 L-BFGS iterations per ADMM iteration, the NRMSE approaches its final level within approximately 10 ADMM iterations. The proposed method requires 13 s per slice, compared with 29 s for DPS using 1000 sampling steps, corresponding to approximately a 2.2-fold speedup. 

The proposed formulation leads to a nonconvex optimization problem. To assess the sensitivity of the proposed algorithm to initialization, we compared four strategies: zero initialization, Gaussian random initialization, uniform random initialization, and ML-based initialization. For ML-based initialization, EM updates were first used to maximize the Poisson log-likelihood and obtain an image estimate $\mathbf{x}_{\text{EM}}$. This estimate was then mapped to the latent space by solving 
\begin{equation}
	\mathbf{x}_0^{0} =
	\arg\min_{\mathbf{x}_0}
	\left\|
	\mathbf{G}_{\boldsymbol{\uptheta}}(\mathbf{x}_0)
	-
	\mathbf{x}_{\text{EM}}
	\right\|_2^2 .
	\label{eq:init_projection}
\end{equation}
As shown in Fig.~\ref{log-likelihood}, the four strategies exhibit different initial NRMSE values but approach nearly identical final NRMSE values, indicating limited sensitivity to initialization in this experiment.

The number of Euler steps controls the numerical accuracy and computational cost of each flow-matching generator evaluation. We examined its effect on reconstruction performance while keeping the other algorithm settings fixed. As shown in Fig.~\ref{Euler}, the NRMSE decreases sharply from 16.5\% with one Euler step to 7.49\% with 10 steps, with only marginal improvements thereafter. These results indicate that 10 Euler steps provide a favorable balance between reconstruction accuracy and the computational cost of evaluating the generator.

Fig.~\ref{real_data} compares all methods on clinical-image-based simulated 3D data at the 25\% dose level, showing representative axial, coronal, and sagittal views with corresponding relative error maps and quantitative metrics. Both DPS and the proposed method use five consecutive axial slices as network input, with a sliding window applied along the axial direction to incorporate information from neighboring slices. ML exhibits substantial noise, while TV suppresses noise but blurs fine anatomical structures. The learning-based methods further improve reconstruction quality, with FBSEM performing competitively with DPS and both outperforming DIP in the reported metrics. The proposed method achieves the best quantitative results in all three views, consistent with the reduced errors visible in the corresponding error maps.

Table~\ref{tab:patient_quant} summarizes the quantitative results for 20 subjects across five dose levels. The proposed method achieves the lowest NRMSE and the highest PSNR and SSIM at every evaluated dose level, supporting its effectiveness on clinical-image-based simulated data.

To evaluate cross-scanner generalization, we compared DPS and the proposed method under two settings: matched training and testing data (Siemens training / Siemens testing) and mismatched training and testing data (United Imaging training / Siemens testing). As shown in Fig.~\ref{cross_sanner}, the proposed method achieves better reconstruction quality than DPS in all three views under both settings. While both methods exhibit reduced reconstruction accuracy under scanner mismatch, the degradation is smaller for the proposed method. These results suggest that the proposed method is more robust than DPS to the training–testing distribution shift in this set of cross-scanner simulation experiments.

\subsubsection{In vivo study}
To evaluate performance on in vivo PET data, we applied the proposed method to 10\% count-level sinograms from an independent patient dataset. The flow-matching prior was trained exclusively on the Siemens challenge images, without retraining on this dataset. As shown in Fig.~\ref{real}, the proposed method suppresses low-count noise while preserving cortical and subcortical structures. Compared with ML and TV, it provides a better balance between noise suppression and structural preservation. Its reconstructions also show closer visual agreement with the full-count reference than those obtained using DIP and DPS.

Fig.~\ref{cov} further evaluates the trade-off between mean caudate activity and the coefficient of variation (CoV) in white-matter regions using 10\% count-level data from 15 subjects. The proposed method achieves mean caudate activity close to the full-count reference while maintaining lower background CoV than the competing methods at comparable activity levels.
\section{CONCLUSION}
\normalcolor
In this paper, we presented an unsupervised PET reconstruction method that uses a pretrained flow-matching generative model as a learned manifold prior. We formulated reconstruction as a regularized Poisson likelihood estimation problem constrained to this generative manifold and developed an ADMM-based algorithm integrating EM-type data-consistency updates with gradient-based latent-space optimization for manifold projection. Experiments on simulated and in vivo PET data demonstrated improved reconstruction performance compared with the evaluated conventional, supervised, and unsupervised learning-based baselines.

Establishing rigorous convergence guarantees for the proposed algorithm is an important direction for future work. Another promising direction is to extend the current optimization-based framework, which yields a point estimate, to sampling-based Bayesian inference for uncertainty quantification, building on our preliminary work \cite{Ran2026_ISBI, Ran2026_IVMSP}.

\bibliographystyle{IEEEtran}
\bibliography{bibliography}

\ifCLASSOPTIONcaptionsoff
  \newpage
\fi

\end{document}